\documentclass[submission,copyright,creativecommons]{eptcs}
\providecommand{\event}{SCSS 2012} % Name of the event you are submitting to
\usepackage{breakurl}             % Not needed if you use pdflatex only.

\usepackage{amssymb}
\usepackage{epsfig}
\usepackage{multirow}
\usepackage{amsmath}
\usepackage{algorithm}
\usepackage{algorithmic}
\usepackage{dsfont}
\usepackage{cite}      % Written by Donald Arseneau
\usepackage{graphicx}  % Written by David Carlisle and Sebastian Rahtz
\usepackage{psfrag}    % Written by Craig Barratt, Michael C. Grant,
\usepackage{subfigure} % Written by Steven Douglas Cochran
\usepackage{url}       % Written by Donald Arseneau
\usepackage{array}
\usepackage{multirow}
\usepackage{amssymb}
\usepackage{alltt}
\usepackage[english]{babel}
\usepackage[titletoc]{appendix}
\usepackage{listings}

\title{On the Verification of a WiMax Design Using Symbolic Simulation}
\author{Salim Ismail Al-Akhras, Sofi\`{e}ne Tahar
\institute{ECE Department, Concordia University, \\ Montreal, QC, Canada}
\email{\{s\_alakhr,tahar\}@encs.concordia.ca}
\and
Gabriela Nicolescu
\institute{CSE Department, Ecole Polytechnique de Montreal,\\ Montreal, QC, Canada}
\email{gabriela.nicolescu@polymtl.ca }
\and
Michel Langevin, Pierre Paulin
\institute{STMicroelectronics Inc., Ottawa, ON, Canada}
\email{\{michel.langevin,pierre.paulin\}@st.com }
}
\def\titlerunning{On the Verification of a WiMax Design Using Symbolic Simulation}
\def\authorrunning{S. I. Al-Akhras, S. Tahar, G. Nicolescu, M. Langevin \& P. Paulin}
\def\copyrightholders{Al-Akhras, Tahar, Nicolescu, Langevin \& Paulin}
\begin{document}
\maketitle

\begin{abstract}
In top-down multi-level design methodologies, design descriptions at higher levels of abstraction are incrementally refined to the final realizations. Simulation based techniques have traditionally been used to verify that such model refinements do not change the design functionality. Unfortunately, with computer simulations it is not possible to completely check that a design transformation is correct in a reasonable amount of time, as the number of test patterns required to do so increase exponentially with the number of system state variables. In this paper, we propose a methodology for the verification of conformance of models generated at higher levels of abstraction in the design process to the design specifications. We model the system behavior using sequence of recurrence equations. We then use symbolic simulation together with equivalence checking and property checking techniques for design verification. Using our proposed method, we have verified the equivalence of three WiMax system models at different levels of design abstraction, and the correctness of various system properties on those models. Our symbolic modeling and verification experiments show that the proposed verification methodology provides performance advantage over its numerical counterpart.
\end{abstract}

\renewcommand*{\arraystretch}{1.5}

\section{Introduction}
\label{intro_sec}
System verification at earlier design stages is extremely important as the cost of fixing bugs at later design stages is usually very high. At higher levels of abstraction, the function of state-of-the-art SoCs and embedded systems is described using programming languages such as C/C++. High level design support is still in its infancy and there is a real need for the development of efficient verification techniques at these higher levels. Traditional simulation based techniques compare simulated design outputs against expected outputs and determine if the design functionality is correct. Since, the number of test patterns required to exhaustively check a design increases exponentially with the number of system state variables, it is infeasible to verify an overall design using exhaustive simulations. Even with carefully selected test vectors designed to cover typical and corner cases, it is often impossible to rule out the presence of design bugs using simulations.

In this paper, we propose an efficient semi-formal verification method for the verification of designs at higher levels of abstraction. We use sequence of recurrence equations (SRE) \cite{1} to mathematically model the system at various levels of abstraction. Then, we use symbolic simulations for the evaluation of design behavior in which multiple input values are encoded as symbols. Finally, we use the results of symbolic simulation on SRE system models along with an equivalence checking technique and an assertion based verification technique to address the verification problem. The use of formal methods, which are exhaustive by nature, guarantees 100\% functional coverage of the design, and symbolic simulations provides scalability. We apply this methodology to an industrial design - STMicroelectronics WiMax modem.

The rest of the paper is organized as follow. We first present related work in Section 2. We then present a brief introduction to Symbolic Simulation and Sequence of Recurrence Equations in Section 3. We describe our modeling and verification methodology in Section 4. Using our proposed methodology, we then verify the STMicroelectronics WiMax modem models in Section 4. We also present experimental results in this section. Finally, Section 5 concludes the paper and discusses possible directions for future work.

\section{Related Work}
Complex telecommunication hardware is being designed using top down multilevel design approaches. In \cite{23}, Deb \emph{et al} propose a Transaction Level Modeling (TLM) based design methodology for refining C and MATLAB functional models of a DSP system into a realistic implementation. In \cite{8}, Fujita \emph{et al} present a similar multi-level system design methodology which uses C/C++, SpecC \cite{72} and SystemC \cite{66} for describing the system. \cite{56} describes a framework for refining functional descriptions to FPGA implementations and presents a physical layer implementation of a WiMax modem. In \cite{38} and \cite{39} STMicroelectronics presents a framework for DSP system design using a set of design refinements applied to C/C++ descriptions for a target system. In this paper, we use the WiMax modem implementation described in \cite{38,39}. \cite{60} presents a WiMax design and verification kit for WiMax certified products. Chiang \emph{et al} used SystemVerilog to validate the physical access layer of WiMax systems \cite{59}. Both \cite{59,60} use computer simulations for the verification and suffer from numerical inaccuracies. Moreover, the absence of bugs cannot be guaranteed unless exhaustive simulations are performed. The verification of the designs at high levels of abstraction, such as the WiMax model, is still an open problem. In this paper, we focus on the verification of WiMax design at the function and the architecture level.

Matsumoto \emph{et al} \cite{4} present an equivalence checking method for two C descriptions using symbolic simulation and prove the equivalence of all variables in the descriptions. We use a similar concept in our proposed methodology to improve the comparison performance of the high level descriptions of hardware designs. In \cite{22}, Abdi \emph{et al} describe a verification method based on model algebra. Systems are described as model algebra expressions. Equivalence of models is checked by proving correctness of model transformation based on a set of predefined rules. If these rules are not used in model transformation, then the correctness of model transformations cannot be proven. Moreover, this work focuses on the correctness of the transitions rather than the functional correctness of the transformed models themselves. In this paper, we propose a higher level symbolic simulation based technique that uses sequence of recurrence equations (SRE) and pattern matching. In \cite{1}, the notion of recurrence equation is extended to describe digital systems for formal verification purposes including support for mathematical reasoning based on symbolic algebra and recurrence equations.

In \cite{33} and \cite{34}, Zaki \emph{et al} used symbolic simulation and SRE to verify properties of continuous Analog and Mixed Signal (AMS) systems. They show that the speed of verification and the coverage of the verification can be enhanced using their method. This work in the AMS field is the inspiration of our work in the system level verification of digital systems.

\section{Preliminaries}
\subsection{Sequence of Recurrence Equations $(SRE)$ }
A recurrence equation or a difference equation is the discrete version of an analogue differential equation. In conventional system analysis, recurrence equations are used in the definition of relations between consecutive elements of a sequence. In \cite{1}, the notion of recurrence equation is extended to describe digital circuits using the normal form: \emph{generalized If-formula}.\\

\noindent \textbf{Definition: Generalized If-formula}
In the context of symbolic expressions, the generalized If-formula is a class of expressions that extend recurrence equations to describe digital systems. Let \emph{K} be a numerical domain $(\mathbb{N}, \mathbb{Z}, \mathbb{Q}, \mathbb{R}$ and $\mathbb{B})$, a generalized If-formula is one of the following: \begin{itemize}
  \item A variable \( X_{i}\)\emph{(n)} or a constant \(C \in K \)
  \item Any arithmetical operation \(\alpha \in \{+,-,\times,\div\}\) between variables \(X_{i}(n) \in K\)
  \item A logical formula: any expression constructed using a set of variables \(X_{i}(n) \in B\) and logical operators: \emph{not}, \emph{and}, \emph{or}, \emph{xor}, \emph{nor} . . . etc.
  \item A comparison formula: any expression constructed using a set of \(X_{i}(n) \in K\) and comparison operator \(\alpha \in \{=, <>, <, =, >, =\}\).
  \item An expression \emph{IF(X, Y, Z)}, where X is a logical formula or a comparison formula and \emph{Y}, \emph{Z} are any generalized If-formula. Here, \emph{IF(x, y, z)}: \( B \times K \times K \rightarrow K \) satisfies the axioms:

\begin{tabular}{ll}
~~~~~ & \emph{IF(True~,~X,~Y)~=~X} \\
~~~~~ & \emph{IF(False,~X,~Y)~=~Y} \\
\end{tabular}
\end{itemize}
\noindent \textbf{Definition: A System of Recurrence Equations (SRE)}
Consider a set of variables \( X_{i}(n) \in K, i \in V = 1 . . . k, n \in Z \), an SRE is a system of the form: \[ X_{i}(n) = f_{i}(X_{j}(n - \gamma)), ( j, \gamma) \in \epsilon_{i}, \forall n \in Z \] where \( f_{i}(X_{j}(n - \gamma)) \) is a generalized If-formula. The set \( \epsilon_{i} \) is a finite non empty subset of \( 1 . . . k \times N \). The integer \( \gamma \) is called the delay.

\subsection{Symbolic Simulation}

Symbolic simulation is a form of simulation where many possible executions of a system are considered simultaneously. The symbolic simulation described in this section relies on rewriting rules based on the algorithms developed in \cite{1} for digital systems. In the context of functional programming and symbolic expressions, we define the following functions.\\

\noindent \textbf{Definition: Substitution.}
Let \emph{u} and \emph{t} be two distinct terms, and \emph{x} a variable. We call $x \rightarrow t$ a substitution rule. We use $Replace(u,x\rightarrow t)$, read ”replace in \emph{u} any occurrence of \emph{x} by \emph{t}”, to apply the rule $x \rightarrow t$ on the expression \emph{u}.
The function \emph{Replace} can be generalized to include a list of rules. \emph{ReplaceList} takes as arguments an expression \emph{expr} and a list of substitution rules $\Re = \{\Re_{1},\Re_{2},...,\Re_{n}\}$. It applies each rule sequentially on the expression. The symbolic simulation function $ReplaceRepeated(Expr;\Re)$ shown in the definition below is based on rewriting by repetitive substitution, which applies recursively a set of rewriting rules $\Re$ on an expression $Expr$ until a fixpoint is reached.\\

\noindent \textbf{Definition: Repetitive Substitution.}
$ReplaceRepeated(expr;\Re)$ applies a set of rules $\Re$ on an expression $expr$ until a fixpoint is reached as shown in the next definition.\\

\noindent \textbf{Definition: Substitution Fixpoint.}
A substitution fixpoint $FP(expr;\Re)$ is obtained, if: $Replace(expr,\Re) \equiv Replace(Replace(expr,\Re ),\Re )$

\noindent Depending on the type of expressions, we distinguish the following kinds of rewriting
rules: \begin{itemize}
  \item \emph{Polynomial Symbolic Expressions} $R_{Math}$: are rules intended for the simplification of polynomial expressions $(\mathbb{R}^{n}[x])$.
  \item \emph{Logical Symbolic Expressions} $R_{Logic}$: are rules intended for the simplification of Boolean expressions and to eliminate obvious ones like $(and(a,a)\rightarrow a)$ and $(not(not(a))\rightarrow a)$.
  \item \emph{If-formula Expressions} $R_{IF}$: are rules intended for the simplification of computations over If-formulae. The definition and properties of the \emph{IF} function, like reduction and distribution, are defined as follows: \\
\begin{tabular}{ll}
  ~~~~~ & \emph{IF Reduction:} ~~~$IF(x;y;y) \rightarrow y$ \\
  ~~~~~ & \emph{IF Distribution:} $f(A1,...,IF(x,y,z),...,An) \rightarrow$\\
  ~~~~~ & ~~~~~~~~~~~~~~~~~~~~~~~~~~$IF(x,f(A1,...,y,...,An),f(A1,...,z,...,An))$ \\
\end{tabular}
  \end{itemize}

\subsubsection{Symbolic Simulation Algorithm.}
The symbolic simulation algorithm used in the symbolic trace computation step is based on rewriting by substitution. The idea is to compute the symbolic execution trace of the SRE model after \emph{n} simulation cycles. During each cycle, the symbolic expressions of each design object are computed using a set of simplification rules. This algorithm is based on repeated substitutions as defined in Algorithm \ref{Algorithm : Repetitive Substitution}. The algorithm repeatedly applies a set of substitution rules \emph{R}, until a fixed point is reached.
%Repetitive Substitution is defined using the following procedure:
\begin{algorithm}
\caption{Repetitive Substitution}
\label{Algorithm : Repetitive Substitution}
\begin{algorithmic}[1]
\STATE $ReplaceRepeated(Expr;\Re)$;
\STATE $Begin$
\REPEAT
\STATE $expr_{1}=ReplaceList(expr,\Re)$
\STATE $expr=expr_{1}$
\UNTIL $FP(expr_{1},\Re)$
\STATE $End$
\end{algorithmic}
\end{algorithm}

Three kinds of symbolic expressions are considered: Algebraic, Logical and If-formula expressions. Each kind is associated with a set of rewriting rules: \(R_{Math}\), \(R_{Logic}\) and \(R_{IF}\).

\begin{itemize}
  \item \textbf{Algebraic expressions \(R_{Math}\):} are Mathematica built-in rules intended for the simplification of polynomial expressions (\(R^{n}[x]\)).
  \item \textbf{Logical symbolic expressions \(R_{Logic}\):} are rules intended for the simplification of Boolean expressions and to eliminate obvious ones like (and(a,a)\(\rightarrow\) a) and (not(not(a)) \(\rightarrow\) a).
  \item \textbf{If-formula expressions \(R_{IF}\) : }are rules intended for the simplification of computations over If-formulas. The definition and properties of the IF function, like reduction and distribution, are used.
\end{itemize}

We add to these rules the trace of the equation at time \emph{n-1} that we consider as rewriting rules of the time (\emph{n-1}) see \cite{1} for more details.

\subsubsection{Verification of Symbolic Traces}
The result of the symbolic simulation is a set of expressions that represent the symbolic trace of the system after \emph{n} cycles. The comparison of expressions is achieved using: \emph{Pattern Matching} \cite{76} and \emph{Equational Theorem Proving} \cite{77}. Pattern matching is used to check that expressions have the desired structure, to find relevant structure, and to substitute the matching part with other expressions. In Mathematica, it is presented as of a regular expression language (Mathematica pattern language) and a set of matching functions. The designer writes properties of the form: \(P\) = \(verify\) (\(U_{i}\),\(S_{i}\)(\(t_{n}\))) where \(U_{i}\) is a regular expression that describes the expected symbolic expression of a simulated object. \(S_{i}(t_{n})\) is the symbolic simulation result of the element \(S_{i}\) after \(t_{n}\) simulation cycles.

\section{Proposed Verification Methodology}
\begin{figure}[!htb]
\centering
\includegraphics[width=6.25in]{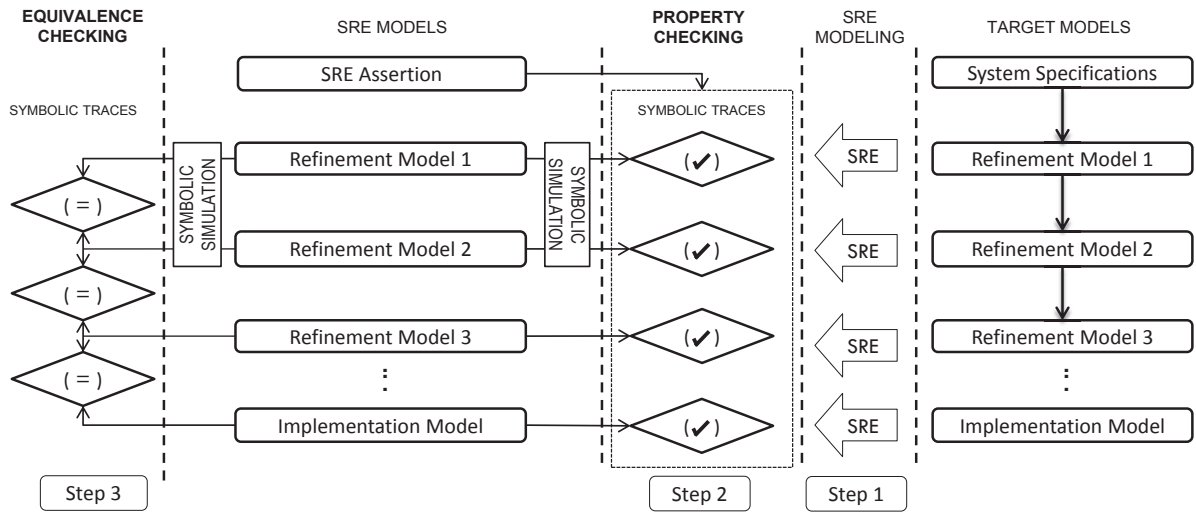}
\caption{\small{Proposed Verification Methodology Framework.}}
\label{Figure: Proposed Verification Methodology Framework000}
\end{figure}

Figure \ref{Figure: Proposed Verification Methodology Framework000} shows the proposed methodology. First, key system specifications (properties) and the design model at each level of abstraction  is translated into a Sequence of Recurrence Equations (SRE)s. Then, two complementary verification processes are applied to these models. The first process uses symbolic simulation to verify the conformance of SRE models to key features of the system. The second process proves the functional equivalence of SRE models. These two processes are incrementally used to verify the correctness of design refinements, and thereby guaranteeing the verification coverage of the design refinements at each level of abstraction and at corner specification points in the design.

Our methodology aims to prove that a system description satisfies a set of properties using pattern matching and equation solving in Symbolic Algebra. This is achieved via several steps as shown in Figure \ref{Figure: Proposed Verification Methodology Framework000}. The system is described using recurrence equations. The properties (or assertions) are algebraic relations between signals of the system. The system description and properties are input to a symbolic simulator that performs a set of transformations by using rewriting rules in order to obtain the symbolic traces. The next step is to use Pattern Matching and Equation solving in Symbolic Algebra to prove the conformance of the properties with the specific system description under test. If the proof is obtained, then the property is verified. Otherwise, we provide counterexamples for the non-proved properties.

\subsection{Computational Equivalence Checking Algorithm}
Algorithm \ref{Algorithm : Computational Equivalence Checking} presents our Computational Equivalence checking algorithm, which we explain in the sequel.

\begin{algorithm}
\caption{Computational Equivalence Checking}
\label{Algorithm : Computational Equivalence Checking}
\begin{algorithmic}[1]

\STATE t = \(t_{0}\);
\STATE \( \phi(t_{0}) = \{ Spec_{j}(t_{0}) \}\) \( 0 < j \leq m ;\)
\WHILE{ \( t \leq K_{Spec}\) }
\STATE \( \phi(t) = SymSim\_Step(\phi) \)
\STATE If NoDeltaCycle then t = t+1
\ENDWHILE
\STATE SPEC = \( \phi(t_{0}+K_{Spec})\)
\STATE t = \(t_{0}\);
\STATE   \( \varphi(t_{0}) =  \{ Imp_{i}(t_{0})\}\)  \(0 < i \leq m;\)
\WHILE{ \( t \leq K_{Imp} \) }
\STATE \( \varphi(t) = SymSim\_Step(\varphi) \)
\STATE If NoDeltaCycle then t = t+1
\ENDWHILE
\STATE IMPL = ReplaceRepeated ( \( \varphi (t_{0}+K_{Imp}) , R_{Abst}  \) )
\STATE MatchQ  \( (\varphi(T) , \phi(T)); \) // T = \(t_{0}\) + k

\end{algorithmic}
\end{algorithm}

\subsubsection{Computing the Trace of the SPEC}

(Lines 1-7): Line 1 first initializes the simulation time \emph{t} to \(t_{0}\) (equal to zero in most cases). The purpose of line 2 is to store the initial SRE of the SPEC model in the variable \(\phi(t_{0})\). Lines 3-6 repeatedly execute a symbolic simulation for \(K_{Spec}\) steps using the symbolic simulation algorithm; the time is advanced only if no more delta cycles are needed. \(K_{Spec}\) is determined by the verifier and it depends on the temporal complexity of the SRE description of the system. For the WiMax application we set \(K_{Spec}\) to 1 because the SRE describing the system is of first order. The variable SPEC stores the computed expressions in line 7. This is equivalent to a new SRE where the time variable is changed to \(T = t_{0}+K_{Spec}\). This traced SRE will be used to compare the traces in line 15.

\subsubsection{Computing the Trace of the IMPL}

(Lines 8-14): In the same way, the trace of the IMPL model is computed using a symbolic simulation for \(K_{Imp}\) steps (same as \(K_{Spec}\)). \(K_{Spec}\) and \(K_{Imp}\) are the number of times (steps), that the symbolic simulation algorithm is repeated on the specification and implementation system descriptions, respectively, to generate the symbolic traces. The verifier should choose \(K_{Spec}\) and \(K_{Imp}\) depending on the temporal complexity of the SRE equations used to describe the system at those levels. So, in general, \(K_{Spec}\) may differ from \(K_{Imp}\), however, in the case of the WiMax modem, we chose \(K_{Spec}\) and \(K_{Imp}\) to be 1, because, the SREs used in the experiment are of first order. The new SRE where the time variable is changed to \(T = t_{0}+K_{Imp}\) is stored to be used to compare the trace of the IMPL model in line 15. In fact, as the IMPL model is more detailed, the direct comparison is not correct. Thus, we need to add some abstraction rules to refine the computed expressions before comparing the results with SPEC. These rewriting rules \(R_{Abst}\) are intended to eliminate calls for functions that convert to integers and rename signals in the IMPL model by their correspondent in SPEC. In line 14, these abstracted expressions are stored in the variable IMPL.

\subsubsection{Comparing Both Traces}

(Line 15): Using pattern matching and algebraic verification, we verify that symbolic expressions in SPEC can be substituted by variables computed in IMPL. The traced symbolic expressions are put in a normal form, and then verified using the function \emph{MatchQ}. This is a built-in function in the computer algebra system, Mathematica 6.0 \cite{6} and it implements the \emph{Pattern Matching} and the \emph{Equational Theorem Proving}. \emph{k} in (line 15) is the maximum of \(K_{Spec}\) and \(K_{Imp}\) and is used as an input to the pattern matching function. For the WiMax experiment, we set \emph{k} = 1. If the verification returns true, then computational equivalence is checked. Otherwise, the pattern matcher gives the non equivalent patterns.

\subsection{Property Checking Algorithm}
Algorithm \ref{Algorithm : Property Checking} presents our proposed property checking Algorithm, which we describe in the following.

\begin{algorithm}
\caption{Property Checking}
\label{Algorithm : Property Checking}
\begin{algorithmic}[1]
\STATE PROP = \{ Prop(IMPL) \};
\STATE t = \(t_{0}\);
\STATE   \( \varphi(t_{0}) =  \{ Imp_{i}(t_{0})\}\)  \(0 < i \leq m;\)
\WHILE{ \( t \leq K_{Imp} \) }
\STATE \( \varphi(t) = SymSim\_Step(\varphi) \)
\STATE If NoDeltaCycle then t = t+1
\ENDWHILE
\STATE IMPL = ReplaceRepeated ( \( \varphi (t_{0}+K_{Imp}) , R_{Abst} \) )
\STATE MatchQ  (IMPL, PROP) // T = \(t_{0}\) + k

\end{algorithmic}
\end{algorithm}

\subsubsection{Storing System Properties}

(Line 1): \emph{Prop (IMPL)} is the set of properties of the system that we want to verify. Those properties are written manually as a system of recurrence equations (SRE).

\subsubsection{Computing the Trace of IMPL}

(Lines 2-8): Similar to what we have done in the equivalence checking part, the trace of the IMPL model is computed using a symbolic simulation for \(K_{Imp}\) steps. In line 7 the new SRE where the time variable is changed to \(T = t_{0}+K_{Imp}\) is stored in IMPL to be used for property checking later. In fact, as the IMPL model is more detailed, the direct property checking is not correct. Thus, we need to add some abstraction rules to refine the computed expressions before comparing the results with PROP. These rewriting rules \(R_{Abst}\) are intended to eliminate calls for functions that convert to integers and to rename signals in the IMPL model by their correspondent ones in PROP. In line 8, these abstracted expressions are stored in the variable IMPL.

\subsubsection{Comparing PROP and IMPL}

(Line 9): Using pattern matching and algebraic verification, we verify that symbolic expressions in PROP can be substituted by variables computed in IMPL. The traced symbolic expressions are put in a normal form, and then verified using the function \emph{MatchQ}. If the verification returns True, then properties are checked. Otherwise, the pattern matcher gives a counterexample.

\section{Modeling and Verification of a WiMax Modem}

\subsection{ST WiMax Modem Models}  \label{subsection_model}
STMicroelectronics provided us with three different C/C++ models of their proposed design each at a different level of abstraction. They are:
\begin{itemize}
  \item Model 1: Functional Level Model.
  \item Model 2: FIFO Based Process Transfer Model.
  \item Model 3: FIFO and Scheduler Based Process Transfer Model.
\end{itemize}

\begin{figure}[!htb]
\centering
\includegraphics[width=6in]{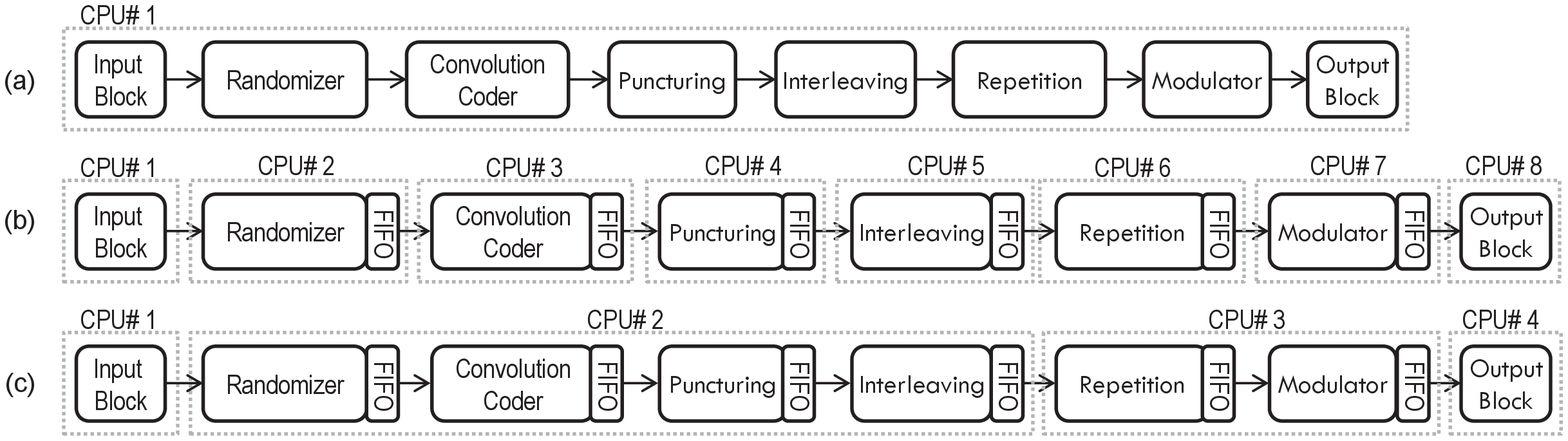}
\caption{\small{(a) Functional Level Model (FL), (b) FIFO Based Process Transfer Model (PTL-8), (c) FIFO and Scheduler Based Process Transfer Model (PTL-4).}}
\label{Figure_WiMax_Models}
\end{figure}

Figure~\ref{Figure_WiMax_Models} shows the three models. The functional model consists of serially connected functional blocks without any additional communication components. These components include the Input block, the Randomizer, the Convolutional coder, the Puncturing block, the Interleaving block, the Repetition block, the Modulator and the Output block. These functional blocks implement various functions specified in the WiMAX standard ~\cite{14}. In the FIFO based process transfer model, a FIFO is used between each of the functional blocks of the system. It allows handling of different timing requirements of the system blocks.  Moreover, in this model, each functional block is mapped to a separate processing unit. All SRE system component models including the FIFO model were extracted from the corresponding C/C++ models provided by STMicroelectronics. Finally, in the FIFO and Schedular based process transfer model, the functionality of more than one functional block were mapped to a single processing unit. In our SRE model, we used a generic scheduler model provided by STMicroelectronics. The scheduler scans the functional blocks in a round-robin manner with a predefined order of the blocks, implementation details can be found in~\cite{saleem_masters_thesis}

We verified the functional equivalence between Model 1 and Model 2, and between Model 2 and Model 3. We first wrote the SRE description of each model using Mathematica (see Appendix A for a sample C++ code and its equivalent SRE description).Then, we validated the correctness of their basic functionality using sample numerical simulation. Next, we generated symbolic traces for SRE models using symbolic simulation in Mathematica. Finally, we used Pattern Matching on those symbolic traces to verify the functional equivalence of all SRE models of the system (Algorithm \ref{Algorithm : Computational Equivalence Checking}). This equivalence implies the equivalence of the corresponding C/C++ models.

We also verified the conformance of these models to important properties of the WiMax transmitter. We wrote those properties as SREs. Then, we used Pattern Matching and Equation Solving Functions from Mathematica to verify these properties (Algorithm \ref{Algorithm : Property Checking}). The conformance of SRE models to these properties implies the conformance of the corresponding C/C++ models to the same properties.

\subsubsection{Single and Multiple Control Scenarios.}

 The physical layer implementation of the ST WiMax modem has various  control signals. These control signals define the behavior of the internal blocks of the WiMax physical layer. Predefined combinations of these control signals are called modes of operation. The WiMax system supports 52 different modes of operation. In a single control scenario simulation,  we perform symbolic simulation assuming that the control signals have predefined value of mode\_0. Whereas, in the multiple control scenario, we perform symbolic simulation seven times once for each mandatory mode of operation supported by the physical layer implementation of the ST WiMax modem~\cite{39,40}.

We validated the correctness of the basic functionality of our SRE models using numerical simulation and compared the simulation results with the corresponding ST functional models. We simulated each of the models for 100 test vectors with a random selection of operation modes and verified that all SRE model outputs were identical to the original ST models. We also generated symbolic traces for the three SRE models of the WiMax modem for seven operation modes. We call this ``mixed simulation mode'' because it used both symbolic and numerical simulation to generate the symbolic traces. Table \ref{TAB:Table5} shows the time and memory utilization of these experiments, which shows the superiority of symbolic simulation over numerical simulation in terms of time requirements.

\begin{table}[ht]
\centering
\caption{Symbolic Simulation Results.}
\label{TAB:Table5}
\scriptsize
\begin{tabular}{|p{1.6in}|p{0.30in}|p{0.375in}|p{0.30in}|p{0.375in}|p{0.30in}|p{0.375in}|p{0.30in}|p{0.375in}|}\hline
& \multicolumn{4}{c|}{Single Control}  & \multicolumn{4}{c|}{Multiple Control}  \\ \cline{2-9}
& \multicolumn{2}{c|}{Symbolic Sim.}  & \multicolumn{2}{c|}{Numerical Sim.}  & \multicolumn{2}{c|}{Symbolic Sim.}  & \multicolumn{2}{c|}{Numerical Sim.}\\ \cline{2-9}
 \multicolumn{1}{|c|}{Model} & {Time}  & {Memory} & {Time}  & {Memory} &{Time}  &{Memory} & {Time}  & {Memory}\\
 \multicolumn{1}{|c|}{  }         & {~(Sec)} & {~(MB)}   & {~(Sec)} & {~(MB)}   &{~(Sec)} &{~(MB)}   & {~(Sec)} & {~(MB)}  \\ \hline\hline
  C/C++ Functional Model & \multicolumn{1}{c|} {N.A.} & \multicolumn{1}{c|} {N.A.} & \multicolumn{1}{c|} {3.21}  &  \multicolumn{1}{c|} {2.30}   &\multicolumn{1}{c|} {N.A.} & \multicolumn{1}{c|} {N.A.} & \multicolumn{1}{c|} {3.21}  &  \multicolumn{1}{c|} {2.30}\\
  \hline
  SRE Functional Level (FL) & \multicolumn{1}{c|} {0.32} & \multicolumn{1}{c|} {13.60} & \multicolumn{1}{c|} {10.00}  &  \multicolumn{1}{c|} {13.45}   &\multicolumn{1}{c|} {2.05} & \multicolumn{1}{c|} {11.96} & \multicolumn{1}{c|} {10.10}  &  \multicolumn{1}{c|} {10.11}\\
  \hline
  SRE Proc. Trans. Level (PLT-8) & \multicolumn{1}{c|} {5.17} & \multicolumn{1}{c|} {19.59}& \multicolumn{1}{c|} {252.4} &   \multicolumn{1}{c|} {16.54}  &\multicolumn{1}{c|} {33.70} & \multicolumn{1}{c|} {12.91} & \multicolumn{1}{c|} {211.3}  &  \multicolumn{1}{c|} {10.00}\\
  \hline
    SRE Proc. Trans. Level (PLT-4) & \multicolumn{1}{c|} {5.26} & \multicolumn{1}{c|} {28.31}& \multicolumn{1}{c|} {266.4} &   \multicolumn{1}{c|} {25.32}  &\multicolumn{1}{c|} {34.18} & \multicolumn{1}{c|} {13.37} & \multicolumn{1}{c|} {222.3}  &  \multicolumn{1}{c|} {11.44}\\
  \hline
\end{tabular}
\normalsize
\end{table}

\subsubsection{Equivalence Checking.}
We verified the computational equivalence between SRE models at different levels of abstraction. We applied the Pattern Matching techniques on the symbolic traces calculated by symbolic simulation. We used Mathematica Pattern Matching built-in function to compare symbolic traces as described in Algorithm \ref{Algorithm : Computational Equivalence Checking}.% (Given in Appendix A).

\subsubsection{Verified Properties.}

We conducted four equivalence checking experiments to prove the following relations:
\begin{itemize}
  \item Equivalence of Functional Model and FIFO based Process Transfer Model in the Single Control scenario.
  \item Equivalence of FIFO based Process Transfer Model and FIFO and Scheduler based Process Transfer Model in the Single Control scenario.
  \item Equivalence of Functional Model and FIFO based Process Transfer Model in the Multiple Control scenario.
  \item Equivalence of FIFO based Process Transfer Model and FIFO and Scheduler based Process Transfer Model in the Multiple Control scenario.
\end{itemize}
The results show that the SRE models at different levels of abstraction are functionally equivalent. In order to guarantee the basic functionality of our equivalence checking algorithm, we injected one bug in one of the SRE models and re-ran the symbolic simulation and equivalence checking experiment. The bug was injected in the SRE FIFO based Process Transfer Model. We changed the functional description of the mapping block. Then, we ran the equivalence experiments again. Now, the results showed non equivalence between the models and returned the nonequivalent symbols from the model's symbolic trace. By inspecting those symbols we found that they were generated only at three modes of operation (0, 1, or 2). From these results, we concluded that the bug was injected in the mapping block implementation only when its puncturing value equals 1/2. More details are provided in~\cite{saleem_masters_thesis}.

Table \ref{Table_EQ_CHECKING} summarizes the performed equivalence and non equivalence experiments along with their time and memory utilization results. The computation time and memory results in Table \ref{Table_EQ_CHECKING} include the time and memory utilization for both symbolic trace computation and pattern matching. From the results in Table~\ref{Table_EQ_CHECKING}, we conclude the following:
\begin{itemize}
  \item The run time of the experiments is linearly proportional to the number of control scenarios. This is interesting because other techniques have exponential increase in time requirements when we increase the execution paths.
  \item Memory requirements of various experiments are comparable to each other.
  \item Since our verification technique depends on pattern matching, we obtained interesting results in the case of non-equivalence. Both time and memory requirements stayed at the same rank as in the equivalence experiments.
\end{itemize}

\begin{table}[ht]
\centering
\caption{Equivalence Checking Results}
\label{Table_EQ_CHECKING}
\scriptsize
\begin{tabular}{|p{0.9in}|p{0.75in}|p{0.35in}|p{0.4in}|p{0.35in}|p{0.4in}|p{0.75in}|}\hline
& &\multicolumn{2}{c|}{Single Control}  & \multicolumn{2}{c|}{Multiple Control} & \\ \cline{3-6}
\multicolumn{1}{|c|}{Models}& \multicolumn{1}{c|}{Design}  & {~Time} & {Memory} & {~Time} & {Memory} & {~~~~~~Result} \\
& & {~(Sec)} & {~(MB)} & {~(Sec)} & {~(MB)} & {} \\ \hline\hline
FL vs. PTL-8 & \multicolumn{1}{c|} { \multirow{2}{*}{Original} } &\multicolumn{1}{c|} {5.17} & \multicolumn{1}{c|} {19.59} & \multicolumn{1}{c|} {33.70}  &  \multicolumn{1}{c|} {12.91} & Equivalent\\
  \cline{1-1} \cline{3-7} % \hline
PTL-8 vs. TPL-4 & & \multicolumn{1}{c|} {5.26} & \multicolumn{1}{c|} {28.31} & \multicolumn{1}{c|} {34.18}  &  \multicolumn{1}{c|} {13.37} & Equivalent\\
  \hline
FL vs. PTL-8 &  \multicolumn{1}{c|} { \multirow{2}{*}{Injected Bugs}} & \multicolumn{1}{c|} {4.96} & \multicolumn{1}{c|} {20.88}& \multicolumn{1}{c|} {32.10} &   \multicolumn{1}{c|} {14.51}& Not Equivalent\\
  \cline{1-1} \cline{3-7} %  \hline
PTL-8 vs. TPL-4 &  & \multicolumn{1}{c|} {6.21} & \multicolumn{1}{c|} {27.21}& \multicolumn{1}{c|} {30.26} &   \multicolumn{1}{c|} {15.67}& Not Equivalent\\
  \hline
%&  Property 4.3 & \multicolumn{1}{c|} {0.02} & \multicolumn{1}{|c|} {1576}& \multicolumn{1}{c|} {0.63} &  \multicolumn{1}{c|} {192}& violated\\ \hline
\end{tabular}
\normalsize
\end{table}

\subsection{Property Checking.}
To verify the conformance of the WiMax models to important properties of the system and to generate counterexamples for properties that turn out to be false, we use the Property Checking algorithm (Algorithm \ref{Algorithm : Property Checking}), described in Section 4.

\subsubsection{Verified Properties.}

We divided the properties with respect to their scope into three categories: \begin{enumerate}
  \item Global Properties: specify a functionality of the whole system.
  \item Local Properties: specify a functionality of a single block.
  \item Control Properties: specify a functionality of a single control configuration (Code Type in the WiMax case)
\end{enumerate}
We wrote several properties of each of these main categories. Here we describe three of them. \begin{itemize}
  \item P1: Eventually all Input Data Bits will be transmitted
  \item P2: Eventually all Input Data Bits with the positions specified by the randomizer bit list will be flipped.
  \item P3: Eventually the Appropriate Puncturing Function will be applied to all Convolution Coded Data Bits in the same order.
\end{itemize}
Next, we translated those properties into SRE (see Appendix B for a sample property written in a form of SRE). After that we applied our proposed Property Checking algorithm to verify their correctness according to the symbolic traces calculated in the symbolic simulation of the model under test. The results of our experiments show that all tested properties are verified to be true under the three models in both single and multiple control scenarios. This shows the conformance of the corresponding C/C++ models from STMicroelectronics to the verified properties.

We also repeated our experiments after injecting the following bugs into all the SRE models.
\begin{enumerate}
  \item Cut one of the data lines between two of the internal blocks.
  \item Changed the randomizer reference array.
  \item Changed the condition checker at the mapping block that specifies the block behavior when the control scenario changes.
\end{enumerate}
The results of the simulation detected all bugs and returned counterexamples that specified the failed property and print the wrong signal value. Table \ref{Table_PROPERTY_CEHCKING} shows the property checking experiments results, together with their time and memory requirements. The results in Table \ref{Table_PROPERTY_CEHCKING} include the time and memory utilization of both symbolic trace computation and pattern matching used in the property checking process. By looking at these results we conclude the following:
\begin{enumerate}
  \item The run time of the experiments is linearly proportional to the number of control scenarios.
  \item Memory requirements of various experiments are close to each other.
  \item Both time and memory requirements stayed at the same rank in both cases, verified true and verified false cases.
\end{enumerate}

\begin{table}[ht]
\centering
\caption{Property Checking Results (Single Control Scenario)}
\label{Table_PROPERTY_CEHCKING}
\scriptsize
\begin{tabular}{|p{0.35in}|p{0.20in}|p{0.3in}|p{0.4in}|p{0.3in}|p{0.4in}|p{0.3in}|p{0.3in}|p{0.4in}|p{0.3in}|p{0.4in}|p{0.3in}|}\hline

\multicolumn{1}{|c|}{} & \multicolumn{1}{c|}{}  & \multicolumn{4}{c|}{Original}& \multicolumn{1}{c}{} &\multicolumn{4}{|c}{Injected Bugs} &\multicolumn{1}{|c|}{}\\ \cline{3-6} \cline{8-11} %%\cline{3-6} \cline{8-11}

 \multicolumn{1}{|c|}{Model} & \multicolumn{1}{c|}{Prop} & \multicolumn{2}{c|}{Single Control}  & \multicolumn{2}{c}{Mult. Control}&  \multicolumn{1}{|c|}{Result} &\multicolumn{2}{c|}{Single Control}& \multicolumn{2}{c|}{Mult. Control}&\multicolumn{1}{c|}{Result} \\ \cline{3-6} \cline{8-11}

&\multicolumn{1}{c|}{erty}& {Time} & {Memory} & {Time} & {Memory} &   &{Time} & {Memory} & {Time} & {Memory}&{}\\ %\hline\hline
 & \multicolumn{1}{c|}{}& {~(Sec)} & {~(MB)} & {~(Sec)} & {~(MB)} &&{~(Sec)} & {~(MB)} & {~(Sec)} & {~(MB)}& \multicolumn{1}{c|}{}\\ \hline\hline

  \multicolumn{1}{|c|}{}   &\multicolumn{1}{c|}{1}& \multicolumn{1}{c} {3.25} & \multicolumn{1}{|c|} {22.05} & \multicolumn{1}{c|} {25.70}  &  \multicolumn{1}{c|} {20.15} & ~True &\multicolumn{1}{c} {3.00} & \multicolumn{1}{|c|} {22.23}& \multicolumn{1}{c|} {30.71} &   \multicolumn{1}{c|} {22.81}& \multicolumn{1}{c|}{False}\\
  \cline{2-12}
   FL&\multicolumn{1}{c|}{2}& \multicolumn{1}{c} {3.10} & \multicolumn{1}{|c|} {23.10} & \multicolumn{1}{c|} {24.31}  &  \multicolumn{1}{c|} {21.23} & ~True &\multicolumn{1}{c} {3.58} & \multicolumn{1}{|c|} {21.25}& \multicolumn{1}{c|} {25.22} &   \multicolumn{1}{c|} {20.60}& \multicolumn{1}{c|}{False}\\
  \cline{2-12}
   &\multicolumn{1}{c|}{3}& \multicolumn{1}{c} {2.92} & \multicolumn{1}{|c|} {22.65}& \multicolumn{1}{c|} {24.10} &   \multicolumn{1}{c|} {23.10}& ~True &\multicolumn{1}{c} {3.10} & \multicolumn{1}{|c|}{20.36}& \multicolumn{1}{c|} {26.10} &   \multicolumn{1}{c|} {13.80}& \multicolumn{1}{c|}{False}\\
  \hline \hline
   &\multicolumn{1}{c|}{1}& \multicolumn{1}{c} {10.52} & \multicolumn{1}{|c|} {22.81} & \multicolumn{1}{c|} {75.20}  &  \multicolumn{1}{c|} {26.20} & ~True &\multicolumn{1}{c} {10.12} & \multicolumn{1}{|c|} {20.15}& \multicolumn{1}{c|} {76.90} &   \multicolumn{1}{c|} {22.05}& \multicolumn{1}{c|}{False}\\
  \cline{2-12}
   PTL-8&\multicolumn{1}{c|}{2}& \multicolumn{1}{c} {11.10} & \multicolumn{1}{|c|} {20.60} & \multicolumn{1}{c|} {80.12}  &  \multicolumn{1}{c|} {23.28} & ~True &\multicolumn{1}{c} {10.22} & \multicolumn{1}{|c|} {21.23}& \multicolumn{1}{c|} {78.15} &   \multicolumn{1}{c|} {23.10}& \multicolumn{1}{c|}{False}\\
  \cline{2-12}
   &\multicolumn{1}{c|}{3}& \multicolumn{1}{c} {10.30} & \multicolumn{1}{|c|} {13.80}& \multicolumn{1}{c|} {78.58} &   \multicolumn{1}{c|} {16.55}& ~True &\multicolumn{1}{c} {11.11} & \multicolumn{1}{|c|} {23.10}& \multicolumn{1}{c|} {79.80} &   \multicolumn{1}{c|} {22.65}& \multicolumn{1}{c|}{False}\\
  \hline \hline
   &\multicolumn{1}{c|}{1}& \multicolumn{1}{c} {10.16} & \multicolumn{1}{|c|} {23.84} & \multicolumn{1}{c|} {79.95}  &  \multicolumn{1}{c|} {22.23} & ~True &\multicolumn{1}{c} {10.16} & \multicolumn{1}{|c|} {26.23}& \multicolumn{1}{c|} {80.94} &   \multicolumn{1}{c|} {22.81}& \multicolumn{1}{c|}{False}\\
  \cline{2-12}
   PTL-4&\multicolumn{1}{c|}{2}& \multicolumn{1}{c} {10.23} & \multicolumn{1}{|c|} {21.02} & \multicolumn{1}{c|} {78.55}  &  \multicolumn{1}{c|} {21.25} & ~True &\multicolumn{1}{c} {11.22} & \multicolumn{1}{|c|} {23.28}& \multicolumn{1}{c|} {97.78} &   \multicolumn{1}{c|} {20.60}& \multicolumn{1}{c|}{False}\\
  \cline{2-12}
   &\multicolumn{1}{c|}{3}& \multicolumn{1}{c} {10.98} & \multicolumn{1}{|c|} {22.36}& \multicolumn{1}{c|} {72.00} &   \multicolumn{1}{c|} {20.36}& ~True &\multicolumn{1}{c} {11.94} & \multicolumn{1}{|c|} {16.55}& \multicolumn{1}{c|} {71.02} &   \multicolumn{1}{c|} {13.80}& \multicolumn{1}{c|}{False}\\
  \hline
  \end{tabular}
\normalsize
\end{table}

\section{Conclusion}
\label{conclusions_SEC}
In this paper, we proposed a semi-formal verification methodology that uses sequence of recurrence equations as a formalism for modeling and property specification. We used symbolic simulation traces and two proposed algorithms for equivalence checking and property checking for design verification.

We illustrated the effectiveness of our methodology by verifying STMicroelectronics WiMax system designs at three different levels of abstraction, one functional level model and two architectural level models. Our experimental results show that the three models are functionally equivalent and the design refinements are correct. In addition, the results show that all models do conform to the specified properties. We detected manually injected bugs in design models and successfully generated counterexamples leading back to the bugs in the design. The performance measurements show that the proposed symbolic simulation based verification method is more efficient than numerical simulations.

We are currently investigating efficient techniques to reduce symbolic simulation time with multiple control signals. In order to automate the modeling part of our methodology, we plan to define transition rules to translate system descriptions from standard programming languages such as C or C++ to SRE. Finally, we plan to apply this semi-formal verification methodology to other more complex system designs.

\bibliographystyle{eptcs}
\bibliography{scss2012_final_salim}

\vspace{3 mm}
\vspace{3 mm}

\appendix

\section{Appendix: Sample C++ Code described in SRE}

An example of writing recurrence equations description of a C++ code is given in this section. The following C++ is just a sample code that has been written to help illustrating the idea of C++ to SRE conversion.

\vspace{3 mm}

\lstset{language=C++, tabsize=2, basicstyle=\ttfamily, numbers=left, numberstyle=\ttfamily, frame=single, breaklines=true}
\begin{lstlisting}
int main(int argc, char *argv[])
{
	class Intermediate_Signal
	{
		int[32] input, output;
	};
	
	class Randomizer_Block
	{
		Intermediat_Signal blockInput , blockOutput, blockControl;
		
		void randomize()
		{
			if (blockControl == MODE_0)
			{
				blockOutput = blockInput;
			}
			else if (blockControl == MODE_1)
			{
				blockOutput = randFunc_01(blockInput);
			}
			else if (blockControl == MODE_2)
			{
				blockOutput = randFunc_02(blockInput);
			}
			else
			{
				blockOutput = INVALID_DATA;
			}			
		}		
	};

	// The actual main
	... Randomizer_Block WimaxRand = new Randomizer_Block;
	WimaxRandomizer->blockControl = MODE_1;
	WimaxRandomizer->blockInput = PREVIOUS_BLOCK_OUTPUT;
	WimaxRandomizer->blockOutput = NEXT_BLOCK_INPUT;
	WimaxRandomizer.randomize();    ...
}
\end{lstlisting}

\vspace{3 mm}

\noindent The following is the SRE representation of the above C++ code.

\vspace{3 mm}

\lstset{basicstyle=\ttfamily, numbers=left, numberstyle=\ttfamily, frame=single, breaklines=true}
\begin{lstlisting}
RAND_OUT =
IF [ RAND_CTRL = MODE_0, RAND_IN,
   (IF [ RAND_CTRL = MODE_1, randFunc_01(RAND_IN),
       (IF [ RAND_CTRL = MODE_2, randFunc_02(RAND_IN),
           INVALID_DATA]) ]) ];
\end{lstlisting}

\section{Appendix: Sample Property Code in SRE}

An example of writing system properties using recurrence equations in Mathematica is given in the following:

\vspace{3 mm}

\noindent ``\textbf{Property P3:} Eventually the appropriate Puncturing Function will be applied to all Convolution Coded Data Bits in the same order"

\vspace{3 mm}

\lstset{basicstyle=\ttfamily, numbers=left, numberstyle=\ttfamily, frame=single, breaklines=true}
\begin{lstlisting}
If[ CodeRate == WMRATE23,

For [i = 0, i < CycleCounter, i++,
	
	If[ PunctOutput[[1, i*3 + 1]] == CCOutput[[1, i*4 + 1]],
	PuncturedSymbols ++,
	Print["Symbol Not Punctured Properly"];,		
	Print["Symbol Not Punctured Properly"];
	];

	If[ PunctOutput[[1, i*3 + 2]] == CCOutput[[1, i*4 + 2]],
	PuncturedSymbols ++,
	Print["Symbol Not Punctured Properly"];,
	Print["Symbol Not Punctured Properly"];
	];

	If[ PunctOutput[[1, i*3 + 3]] == CCOutput[[1, i*4 + 4]],
	PuncturedSymbols ++,
	Print["Symbol Not Punctured Properly"];,
	Print["Symbol Not Punctured Properly"];
	];
];,

If[ CodeRate == WMRATE46, ... , ...];
];
\end{lstlisting}

%\nocite{*}
%\bibliographystyle{eptcs}
%\bibliography{generic}
\end{document}